\documentclass[12pt]{article}
\usepackage{url}
\usepackage{epsfig,epsf}
\usepackage{graphicx,graphics}
\usepackage{amssymb,amsmath}
\usepackage{verbatim,enumerate}
\usepackage[font=scriptsize]{caption}
\usepackage{verbatim,multicol,color}
\usepackage{multirow}
\usepackage{booktabs}
\usepackage{multicol}
\usepackage[english]{babel}
\usepackage{blindtext}
\usepackage{subfig}
\usepackage{slashbox}
\usepackage{natbib}
\usepackage{float}
\usepackage{mathrsfs}
\RequirePackage[colorlinks,citecolor=blue,urlcolor=blue]{hyperref}

\def\boxit#1{\vbox{\hrule\hbox{\vrule\kern6pt
          \vbox{\kern6pt#1\kern6pt}\kern6pt\vrule}\hrule}}

\def\bse{\begin{eqnarray*}}
\def\ese{\end{eqnarray*}}
\def\be{\begin{eqnarray}}
\def\ee{\end{eqnarray}}
\def\bq{\begin{equation}}
\def\eq{\end{equation}}
\def\bse{\begin{eqnarray*}}
\def\ese{\end{eqnarray*}}

\newcommand{\calX}{{\mathcal{X}}}

\newcommand{\bsy}{\boldsymbol}
\newcommand{\iid}{\stackrel{\mathrm{iid}}{\sim}}

\newcommand{\mbf}{\mathbf}

\begin{document}
\baselineskip=28pt
\begin{center} {\Large{\bf High-order Composite Likelihood Inference for Max-Stable Distributions and Processes}}
\end{center}

\baselineskip=12pt

\begin{center}
\large
Stefano Castruccio{\footnotemark[1]}, Rapha\"el Huser$^2$ and Marc G. Genton{\footnotemark[2]}  
\end{center}

\footnotetext[1]{
\baselineskip=10pt School of Mathematics and Statistics, Newcastle University, Newcastle Upon Tyne NE1 7RU, United Kingdom. E-mail: stefano.castruccio@ncl.ac.uk}

\footnotetext[2]{
\baselineskip=10pt CEMSE Division, King Abdullah University of Science and Technology, Thuwal 23955-6900, Saudi Arabia. E-mail: raphael.huser@kaust.edu.sa, marc.genton@kaust.edu.sa}

\baselineskip=16pt

\begin{center}
{{\bf Abstract}}
\end{center}

In multivariate or spatial extremes, inference for max-stable processes observed at a large collection of locations is a very challenging problem in computational statistics, and current approaches typically rely on less expensive composite likelihoods constructed from small subsets of data. In this work, we explore the limits of modern state-of-the-art computational facilities to perform full likelihood inference and to efficiently evaluate high-order composite likelihoods. With extensive simulations, we assess the loss of information of composite likelihood estimators with respect to a full likelihood approach for some widely-used multivariate or spatial extreme models, we discuss how to choose composite likelihood truncation to improve the efficiency, and we also provide recommendations for practitioners. This article has supplementary material online.

\baselineskip=16pt

\vskip 2mm
\noindent
{\bf Key words:}  {\footnotesize composite likelihood; efficiency; max-stable process; parallel computing; spatial extremes}

\par\medskip\noindent
{\bf Short title}: High-order Composite Likelihoods for Max-Stable Processes

\pagenumbering{arabic}
\baselineskip=26pt

\section{Introduction}

In the environmental statistics community, the development of models and inference methods for multivariate or spatial extremes is a very active area of research \citep{da12b}. The max-stability property that underpins these extremal models justifies, at least asymptotically, extrapolation beyond the temporal scale of the observed data \citep{Davison.etal:2013}. In this framework, widely-used max-stable processes include the \cite{Smith:1990}, \cite{Schlather:2002}, Brown--Resnick \citep{br77,ka09} and extremal~$t$ \citep{Nikoloulopoulos.etal:2009,op13} models. However, their practical use on big data sets has been hampered by the computationally challenging nature of likelihood functions, whose evaluation for frequentist or Bayesian inference requires computing and storing a prohibitively large amount of information. This challenge, arising even with moderate sample sizes, contrasts with that of the classical Gaussian-based geostatistics literature, where, thanks to recent advances, hundreds of thousands of locations, or even millions, can be handled using a wide variety of techniques; see the review by \cite{Sunetal2012} and references therein, as well as recent work such as \cite{Nychkaetal2014} and \cite{SunStein2014}.  
Indeed, a single likelihood evaluation for a max-stable process observed at $Q$ locations requires O$(Q^Q)$ operations and storage, which, compared to the ``Big Data'' problem of fitting a large multivariate normal distribution that requires O$(Q^3)$ operations and O$(Q^2)$ storage, is on a different level of computational complexity.

The cumulative distribution function of a max-stable process $Z(\mbf{x})$ observed at $Q$ locations $\mbf{x}_1,\ldots,\mbf{x}_Q\in \calX\subset\mathbb R^2$ may be written as  
\begin{equation}\label{distrMS}
\mathbb{P}\{Z(\mbf{x}_1)\leq z_1,\ldots,Z(\mbf{x}_Q)\leq z_Q\}=\exp[-V\{T(\mbf{z})\}],
\end{equation}
where $T(\mbf{z})$ is a suitable marginal transformation, $\mbf{z}=(z_1,\ldots,z_Q)^\top$ and $V(\mbf{z})$ is a function called the exponent measure; see, e.g., \citet{Huser.etal:2015}. The corresponding density can be obtained by computing the derivative of (\ref{distrMS}) with respect to all elements of $\mbf{z}$. This results in a combinatorial explosion of terms as a function of $Q$ (e.g., the number of monitoring stations in spatial applications), which requires substantial computational power and a prohibitively large amount of information storage if $Q$ is large (more details in Section \ref{inf_mul}). 

To circumvent these computational challenges, the common approach has been to rely on composite likelihoods \citep{li88,va11}, based on pairs \citep{pa10} or triples \citep{gen11,hu13,sa14}. These misspecified likelihoods enable consistent but less efficient estimation of parameters under mild regularity conditions, and are computationally convenient: the number of terms involved decreases dramatically, and each of these terms is less expensive to compute. Although the loss of information entailed by pairwise or triplewise approaches has been largely investigated in various contexts \citep[see, e.g.,][]{co04,hj08,da11,be12,Eidsvik.etal:2014}, little has been done to assess their performance, and that of higher-order counterparts, in the context of extremes. \citet{Huser.etal:2015} compared several likelihood inference methods for multivariate extremes, but they focused on the multivariate logistic model only, and did not consider composite likelihoods of orders higher than two (except for the full likelihood). \citet{bi14} proposed partition-based composite likelihoods, but they assessed their performance for blocks of maximum dimension five, focusing on a clustered max-stable process and on the \citet{Schlather:2002} process. Furthermore, they did not compare composite likelihoods of different orders, and did not provide any assessment of the associated computational burden. Our paper not only complements the existing literature by conducting an extensive simulation study assessing the practical performance of low- and high-order composite likelihoods for several max-stable models, but it also offers a computational perspective on this problem by exploring the limits of likelihood-based inference, suggesting how state-of-the-art computational resources may be efficiently used in this framework, and giving recommendations to practitioners. We also demonstrate that suitably truncated low-order composite likelihoods can improve the inference, while decreasing the computational time and thus allowing analyses in higher dimensions. Our simulation study focuses on max-stable models of increasing computational complexity: we start with the widely-studied multivariate logistic model \citep{gu60}; then, we consider the model for spatial extremes advocated by \cite{re12}, which can be viewed as a spatial generalization of the logistic model; finally, we investigate a related, likely more realistic, stationary process, namely the Brown--Resnick model.

The rest of the paper is organized as follows: Section \ref{ms_proc} gives some background on multivariate or spatial extremes and introduces the models used in this work. Section \ref{inf_mul} discusses inference for extremes based on full or composite likelihoods and describes the related computational challenges. In Section \ref{sim_stud}, the performance of composite likelihoods is assessed by simulation, and truncation strategies and other likelihood approximations are discussed. Section~\ref{cost} provides estimates of memory and computational requirements in high dimensions, as well as practical recommendations, and Section \ref{concl} concludes with perspectives on inference for extremes with current and future computer architectures.
\section{Extreme-Value Theory and Models}\label{ms_proc}
Suppose that $Y_n$ is a sequence of independent and identically distributed (iid) random variables. Furthermore, let there exist sequences of normalizing constants, $a_n>0$ and $b_n$, such that $Z_n=a_n^{-1}(\max_{i=1,\ldots,n}Y_i-b_n)$ converges in distribution to a non-degenerate random variable $Z$, as $n\to\infty$. Then, the distribution of $Z$ has to be of the form
\begin{equation}\label{ex_un}
\mathbb{P}(Z\leq z)=\exp\left[-\left\{1+\xi\left(\frac{z-\mu}{\sigma}\right)\right\}_+^{-1/\xi} \right],
\end{equation}
where $a_+=\max(0,a)$ and $\mu$, $\sigma>0$ and $\xi$ are denominated the location, scale and shape parameters, respectively. The case $\xi=0$ is formally undefined but is understood as the limit as $\xi\to 0$. The right-hand side of (\ref{ex_un}) is referred to as the generalized extreme-value family, or in short GEV$(\mu,\sigma,\xi)$, and has been extensively used as a model for block maxima; see, e.g., \cite{co01} or the review paper by \cite{Davison.Huser:2015}.

\begin{sloppypar}
Likewise, in the multivariate setting, suppose that $\mbf{Y}_1=(Y_{1;1},\ldots,Y_{1;Q})^\top,\mbf{Y}_2=(Y_{2;1},\ldots,Y_{2;Q})^\top,\ldots$, is a sequence of iid random vectors with unit Fr\'echet margins, i.e., each component is distributed according to (\ref{ex_un}) with $\mu=\sigma=\xi=1$; the general case is easily treated by using the probability integral transform to put data on the unit Fr\'echet scale. Then, as $n\to\infty$, the vector of renormalized componentwise maxima $\mbf{Z}_n=n^{-1}(\max_{i=1,\ldots,n}Y_{i;1},\ldots,\max_{i=1,\ldots,n}Y_{i;Q})^\top$ converges in distribution \citep[see, e.g.,][]{co01} to a random vector $\mbf{Z}$ distributed according to a multivariate extreme-value (or equivalently, max-stable) distribution \eqref{distrMS}, with the marginal transformation $T(\mbf{z})$ being the identity, and where the exponent measure $V(\mbf{z})$ is a positive function such that $V(a^{-1}\mbf{z})=aV(\mbf{z})$ for any $a>0$ and $\mbf{z}>\mbf{0}$ (i.e., $V$ is homogeneous of order $-1$). This function satisfies the marginal constraints $V(z\mbf{e}_q^{-1})=z^{-1}$ for any $z>0$ and $q=1,\ldots,Q$, where $\mbf{e}_q$ denotes the $q$th canonical basis vector in $\mathbb{R}^Q$, and $1/0=\infty$ by convention. Numerous multivariate models have been proposed in the literature \citep[see, e.g.,][]{Joe:1997,Tawn:1988b,ta90,Cooley.etal:2010,Ballani.Schlather:2011,Segers:2012}. 

\begin{figure}[t!]
\centering
\includegraphics[width=13cm,keepaspectratio]{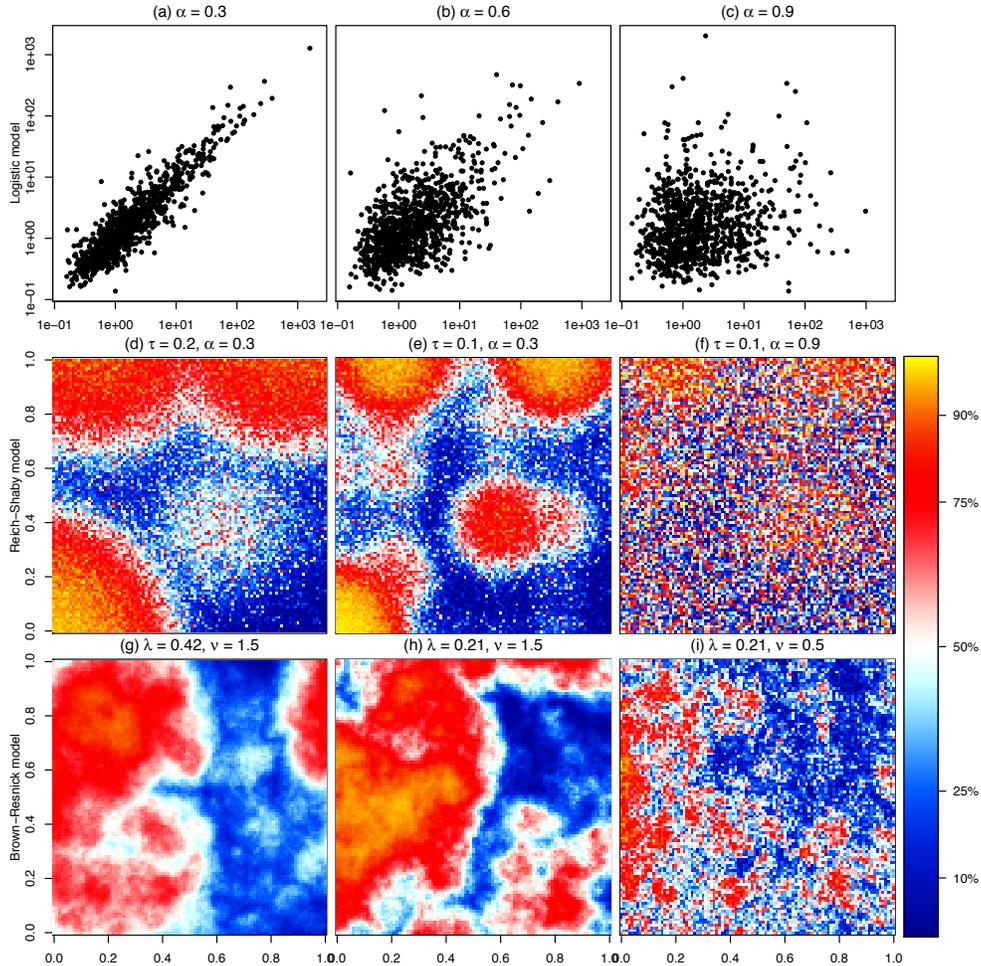}
\caption{Examples of realizations for the two-dimensional logistic model (a-c), Reich--Shaby process with $36$ knots on a regular grid $[0,0.2,\ldots,1]^2$ (d-f), and Brown--Resnick process with semi-variogram $\gamma(\mbf{h})=(\|\mbf{h}\|/\lambda)^{\nu}$ (g-i). The latter are simulated using the {\tt RandomFields} package of the statistical software {\tt R}. Marginal quantiles are indicated by the color scale. Parameters are indicated above the corresponding panels. The grid spacing is 0.005.}
\label{real_plot}
\end{figure}

The first proposed and most widely-used model is the symmetric logistic model \citep{gu60}, which assumes that
\begin{equation}\label{logistic}
V(\mbf{z})=\left(\sum_{q=1}^Q z_q^{-1/\alpha} \right)^\alpha,
\end{equation}
for some dependence parameter $0<\alpha\leq1$. When $\alpha=1$, the components of $\mbf{Z}$ are mutually independent, whereas perfect dependence is attained as $\alpha\to0$; see typical realizations in Figure \ref{real_plot}(a-c). By symmetry, components of $\mbf{Z}$ are equidependent. A possible asymmetric extension of (\ref{logistic}) is to consider a max-mixture of logistic variables. More precisely, for $l=1,\ldots,L$, consider independent random vectors $\mbf{Z}_l=(Z_{l;1},\ldots,Z_{l;Q})^\top$ distributed according to (\ref{distrMS}) and (\ref{logistic}) with dependence parameters $0<\alpha_l\leq 1$. Then, the random vector constructed as $\left(\max_{l=1,\ldots,L} w_lZ_{l;1},\ldots,\max_{l=1,\ldots,L} w_lZ_{l;Q}\right)^\top$, with non-negative weights $w_1,\ldots,w_L$ such that $\sum_{l=1}^Lw_l=1$, is distributed as (\ref{distrMS}) with
\begin{equation}\label{LogisticMixture}
V(\mbf{z})=\sum_{l=1}^L\left\{\sum_{q=1}^Q\left({z_q\over w_l}\right)^{-1/\alpha_l}\right\}^{\alpha_l}.
\end{equation}
This model is different from, but closely related to, the asymmetric logistic model proposed by \citet{ta90}. Although \eqref{LogisticMixture} is more flexible than \eqref{logistic}, the number of parameters equals $2L-1$ and thus increases dramatically for large $L$, and, for practical use, the data structure should be exploited to construct simpler models. 

In the spatial framework, \citet{re12} proposed a max-stable process with finite-dimensional distributions that may be expressed through (\ref{LogisticMixture}). Specifically, let $Z(\mbf{x})$ $(\mbf{x}\in\calX\subset\mathbb R^2)$ be a spatial process defined on the plane as $Z(\mbf{x})=U(\mbf{x})\theta(\mbf{x})$, where $U(\mbf{x})\iid$ GEV$(1,\alpha,\alpha)$ is a random noise process, and $\theta(\mbf{x})=\left\{\sum_{l=1}^L A_l w_l(\mbf{x})^{1/\alpha}\right\}^\alpha$, with $\sum_{l=1}^L w_l(\mbf{x})=1$ and $w_l(\mbf{x})\geq0$ for any $\mbf{x}\in\calX$. The variables $A_l$ are underlying independent random effects distributed according to the $\alpha$-stable distribution ($0<\alpha\leq 1$) (see \citet{st09} for some background on $\alpha$-stable variables), and the deterministic weights $w_l(\mbf{x})$ capture the latent  spatial structure. Then, for any finite subset of locations $\{\mbf{x}_1,\ldots,\mbf{x}_Q\}\subset\calX$, it can be shown that the random vector $\{Z(\mbf{x}_1),\ldots,Z(\mbf{x}_Q)\}^\top$ is distributed according to (\ref{distrMS}) with
\begin{equation}\label{RSmodel}
V(\mbf{z})=\sum_{l=1}^L\left[\sum_{q=1}^Q \left\{{z_q\over w_l(\mbf{x}_q)} \right\}^{-1/\alpha}\right]^{\alpha},
\end{equation}
which is a special case of \eqref{LogisticMixture} with fewer parameters. The weights may be further written as $w_{l}(\mbf{x})=k_{l}(\mbf{x})\left\{\sum_{l=1}^L k_{l}(\mbf{x})\right\}^{-1}$, where $k_{l}(\mbf{x})=\frac{1}{2\pi\tau^2}\exp\left\{-\frac{1}{2\tau^2}(\mbf{x}-\mbf{v}_l)^{\top}(\mbf{x}-\mbf{v}_l)\right\}$, for $\tau>0$, and some fixed knots $\mbf{v}_1,\ldots,\mbf{v}_L\in\calX$. For known knots, this model has only two unknown parameters: $\alpha$ controls the amount of noise in the resulting random field, and $\tau$ is a range parameter. Smaller values of $\alpha$ result in less noisy processes (with $\alpha\to0$ corresponding to deterministic profiles), while increasing $\tau$ implies larger-scale processes; see typical realizations in Figure~\ref{real_plot}(d-f). In particular, it can be shown that, as $\alpha\to0$ and as regularly spaced knots get denser (with $L\to\infty$), model \eqref{RSmodel} converges to the stationary max-stable model proposed by \cite{Smith:1990}. Although appealing, the Reich--Shaby model has several peculiarities. First, the process is stationary and isotropic only as $L\to\infty$. Second, as illustrated in Figure~\ref{real_plot}(e), realizations from \eqref{RSmodel} may be strongly affected by the locations of the knots, if the range parameter $\tau$ is small compared to the knot spacings. And finally, realized random fields are either discontinuous in every point (with $\alpha\in(0,1]$) or analytic almost everywhere (with $\alpha\to0$). Despite these features, which may be seen as limitations in real applications, this model is simple, intuitive, computationally convenient, and inference can be performed at practically the same cost as for the logistic distribution. These computational advantages have proven to be especially useful in Bayesian hierarchical modeling of spatial extremes \citep{re12}.
\end{sloppypar}

A popular and likely more realistic stationary max-stable model is the Brown--Resnick process \citep{br77,ka09}, which may be represented as $Z(\mbf{x})=\sup_{i\in \mathbb{N}} W_i(\mbf{x})/T_i$; the $W_i(\mbf{x})$s are independent copies of a random process $W(\mbf{x})=\exp\{\varepsilon(\mbf{x})-\gamma(\mbf{x})\}$, where $\varepsilon(\mbf{x})$ is an intrinsically stationary Gaussian random field with semi-variogram $\gamma(\mbf{h})$ and $\varepsilon(\mbf{0})=0$ almost surely, and the $T_i$s are occurrences of a unit-rate Poisson process on the positive real line. Such a process has finite-dimensional distributions that may be written as (\ref{distrMS}) with exponent measure
\begin{equation}\label{BRmodel}
V(\mbf{z})=\sum_{q=1}^Q {\Phi_{Q-1}(\bsy{\eta}_q;\mbf{0},\mbf{R}_q)\over z_q},
\end{equation}
where $\Phi_D(\mbf{a};\bsy{\mu},\bsy{\Sigma})$ is the cumulative distribution function of a $D$-variate normal variable with mean $\bsy{\mu}$ and covariance $\bsy{\Sigma}$, evaluated at $\mbf{a}$. Quantities $\bsy{\eta}_q$ and $\mbf{R}_q$ appearing in (\ref{BRmodel}) are defined in the Appendix. The Brown--Resnick process with fractional Brownian semi-variogram, i.e., $\gamma(\mbf{h})=(\|\mbf{h}\|/\lambda)^{\nu}$, $\lambda>0$, $\nu\in(0,2]$, is illustrated in Figure~\ref{real_plot}(g-i). In this case, $\lambda$ is a range parameter, whereas $\nu$ is a smoothness parameter with higher values indicating smoother processes. In particular, the isotropic \citet{Smith:1990} process (with analytical storm profiles) is recovered when $\nu=2$. Hence, both the Reich--Shaby and Brown--Resnick models can be viewed as non-smooth extensions of the Smith model, but they are not nested in each other.

\section{Inference and Computational Challenges}\label{inf_mul}
Likelihood inference for multivariate or spatial extremes is computationally challenging. Although the cumulative distribution function can be easily written as \eqref{distrMS}, the density for a single vector $\mbf{z}=(z_1,\ldots,z_Q)^\top$ is much more complicated since it corresponds to the derivative of \eqref{distrMS} with respect to all components of $\mbf{z}$. Assume that $V(\mbf{z})=V(\mbf{z}\mid \bsy{\theta})$, i.e., the exponent measure is parametrized by some $\bsy{\theta}\in\Theta\subset\mathbb R^{p}$, and that all derivatives of $V(\mbf{z}\mid \bsy{\theta})$ exist. The likelihood in the case of unit Fr\'echet margins, where $T(\mbf{z})$ in \eqref{distrMS} is the identity, is
\begin{equation}\label{ms_density}
L_Q( \bsy{\theta}\mid\mbf{z})=\exp\left\{-V(\mbf{z}\mid \bsy{\theta})\right\}\sum_{\mathcal{P} \in \mathscr{P}_{\mbf{z}}}\prod_{S \in \mathcal{P}}\left\{-V_{S}(\mbf{z}\mid \bsy{\theta})\right\},
\end{equation}
where $\mathscr{P}_{\mbf{z}}$ is the collection of all partitions of $\mbf{z}$ , $S\neq \emptyset$ is a particular set in partition $\mathcal{P}\in\mathscr{P}_{\mbf{z}}$, and $V_{S}(\mbf{z}\mid \bsy{\theta})$ is the partial derivative of $V(\mbf{z}\mid \bsy{\theta})$ with respect to the elements in the set $S$. The main issues regarding the computation of (\ref{ms_density}) are the following:
\begin{enumerate}
\item[1)] Closed-form expressions for $V(\mbf{z}\mid \bsy{\theta})$ and $V_{S}(\mbf{z}\mid \bsy{\theta})$ are not always available. Although Monte Carlo methods could in principle be used to approximate these functions for some models, this would certainly result in a prohibitively large amount of computations, especially in large dimensions. In this paper, we focus on the three models, (\ref{logistic}), (\ref{RSmodel}) and (\ref{BRmodel}), where closed forms are known;
\item[2)] The $2^Q-1$ partial derivatives of $V(\mbf{z}\mid \bsy{\theta})$, corresponding to all non-empty subsets $S$ of $\{1,\ldots,Q\}$ must be computed. Depending on the class of models, this step can be more or less demanding. In this work, the partial derivative evaluation is increasingly complex as we consider models \eqref{logistic}, (\ref{RSmodel}) and (\ref{BRmodel});
\item[3)] The vector of partial derivatives must be assembled to compute the sum across all partitions. As the cardinality of $\mathscr{P}_{\mbf{z}}$ equals $B_Q$, the Bell number of order $Q$ \citep{gr88}, which grows more than exponentially with $Q$, the storage of very large data structures is required. 
\end{enumerate}

Depending on the model considered, 2) and 3) could prevent likelihood evaluation for high-dimensional data, and several solutions can be devised to improve computations. If 2) is the most computationally demanding operation, the evaluation of the $2^Q-1$ derivatives can be performed in parallel across processors. Since in our simulation study (see Section~\ref{sim_stud}), multiple independent experiments are performed, we have found that parallelization across experiments is the most efficient solution, but if the goal is to analyze one data set, this option can greatly reduce the computational time if $Q$ is large. Combining the derivatives into a sum across all partitions in 3) is the main limitation in high dimensions and cannot be easily improved with parallelization. For each likelihood evaluation, the vector $\mbf{D}_Q$ of all the $2^Q-1$ partial derivatives is computed, and in order to efficiently assemble the sum across all partitions, an efficient strategy is to precompute $\mbf{P}_Q$, a $B_Q\times2$ cell array (with elements stored as 16-bit integers instead of double precision), with each row consisting of a different partition. The first column consists of the sets in a partition, and the second consists of the indexes of the sets in $\mbf{D}_Q$. This approach allows for efficient storage, but large $Q$s make it impossible to store $\mbf{P}_Q$. To avoid this, it is possible to divide $\mathscr{P}_{\mbf{z}}$ into groups and to compute the sum dynamically for each group. In our experience, the computation of a larger number of smaller subsets of partitions is computationally very demanding (see Section \ref{cost} for a discussion about groups of Stirling partitions) and, since this needs to be performed for each likelihood evaluation, it has proven to be very slow. In Section~\ref{cost}, we show that a full likelihood evaluation, implemented efficiently on a powerful computer, can be performed in maximum dimension $Q=11$ in reasonable time. Parallelization across experiments implies the creation of a copy of $\mbf{P}_Q$ for each processor but if the goal is the analysis of one data set, a single data structure is created and the computation could possibly be increased to $Q=12$ or $Q=13$ if a large memory is available; computation for higher dimensions is completely out of reach with current technologies.

To avoid these shortcomings, the standard approach \citep{pa10,gen11,hu13,sa14,bi14} is to consider composite likelihoods. If we denote by $\mbf{z}_q$ a vector of $q$ elements from a $Q$-dimensional vector $\mbf{z}$ and $C_{\mbf{z}_q}(Q)$ the collection of all possible sub-vectors $\mbf{z}_q$, then a composite likelihood of order $q$ may be expressed as
\begin{equation}\label{ms_density_composite}
CL_q(\bsy{\theta}\mid\mbf{z})=\prod_{\mbf{z}_q \in C_{\mbf{z}_q}(Q)}L_q(\bsy{\theta}\mid\mbf{z}_q)^{\omega_{\mbf{z}_q}},
\end{equation}
where each contribution $L_q(\bsy{\theta}\mid\mbf{z}_q)$ is a likelihood term of order $q$, as defined in (\ref{ms_density}), and the real numbers $\omega_{\mbf{z}_q}$ are positive weights that do not necessarily sum up to one. In general, evaluation of $L_q(\bsy{\theta}\mid\mbf{z}_q)$ is more convenient than $L_Q(\bsy{\theta}\mid\mbf{z})$ for small $q$s, 
but small $q$s imply large sets $C_{\mbf{z}_q}(Q)$, so the computational time is not necessarily monotonically increasing with $q$, as is shown in Section \ref{cost}. Since composite likelihoods are built from valid likelihood terms, they inherit some of the large-sample properties from the full likelihood. More precisely, suppose that $m$ independent replicates $\mbf{z}_1,\ldots,\mbf{z}_m$ of an extreme-value distributed vector are observed, and consider the full and composite likelihoods constructed from (\ref{ms_density}) and (\ref{ms_density_composite}) as
\begin{equation}\label{likelihoods}
L_Q( \bsy{\theta}\mid\mbf{z}_1,\ldots,\mbf{z}_m)=\prod_{i=1}^m L_Q( \bsy{\theta}\mid\mbf{z}_i),\qquad CL_q( \bsy{\theta}\mid\mbf{z}_1,\ldots,\mbf{z}_m)=\prod_{i=1}^m CL_q( \bsy{\theta}\mid\mbf{z}_i).
\end{equation}
Let $\hat{\bsy{\theta}}$ and $\hat{\bsy{\theta}}_C$ denote the estimators of $\bsy{\theta}$ maximizing the full likelihood and composite likelihood in (\ref{likelihoods}), respectively. Then, under mild conditions, $\hat{\bsy{\theta}}$ and $\hat{\bsy{\theta}}_C$ are both strongly consistent as $m\to\infty$, asymptotically Gaussian, and converge at the same rate, namely $\sqrt{m}$ (see \citealp{pa10}). However, the variability of $\hat{\bsy{\theta}}_C$ is typically larger than that of $\hat{\bsy{\theta}}$, and it depends on the choice of weights $\omega_{\mbf{z}_q}$ in (\ref{ms_density_composite}). When weights are ignored, i.e., $\omega_{\mbf{z}_q}=1$, we refer to the corresponding composite likelihood as a \emph{complete composite likelihood}; when the $\omega_{\mbf{z}_q}$s are not all equal, we use the expression \emph{weighted composite likelihood}; when binary weights are used, i.e., $\omega_{\mbf{z}_q}=0$ or $1$, we refer to it as a \emph{truncated composite likelihood}. The latter is also called tapered composite likelihood by \citet{sa14}. The performance of complete composite likelihoods are studied in Section~\ref{complete_complik} for the models introduced in Section~\ref{ms_proc}, and truncated alternatives are explored in Section~\ref{lik_appr}. As we will show, truncation allows us to drastically reduce the computational time, and in some cases also to improve the efficiency. Section~\ref{cost} contains more detailed discussion on the computational cost of composite likelihoods.

\section{Performance of Composite Likelihoods}\label{sim_stud}
\subsection{General setting}
In this section, we detail simulation studies for the three models mentioned in Section \ref{ms_proc}: we consider them in order of increasing computational complexity, considering first the logistic model (\ref{logistic}), then the Reich--Shaby model (\ref{RSmodel}) and finally the Brown--Resnick model (\ref{BRmodel}). The goal of this section is to assess the improvement of high-order composite likelihoods compared to the traditional pairwise and triplewise approaches, in terms of root mean squared error with respect to the true parameter values. We deliberately choose a relatively small number $Q$ of locations ($Q=11$ for the logistic and the Reich--Shaby models, and $Q=9$ for the Brown--Resnick model), so that full likelihoods can be computed in a reasonable time, and we provide estimated projections of the computational cost in higher dimensions in Section~\ref{cost}.

All simulations (obtained with the same random seed) were performed on a fully dedicated 39-node cluster with 20 cores (and 512 Gb of RAM) per node, for a total of 780 cores. The algorithms were implemented in MATLAB, and each likelihood maximization was performed using a Nelder--Mead algorithm, allowing at most 1000 iterations and a tolerance of convergence of $0.01$ between successive iterates. Given the small dimensionality of the parameter space ($p=1$ or $2$), these specifications were sufficient to achieve accurate results for all experiments. 

\subsection{Complete composite likelihoods}\label{complete_complik}
\vspace{-.2cm}

\begin{figure}[t!]
\centerline{\includegraphics[width=0.53 \linewidth,height=4.6cm]{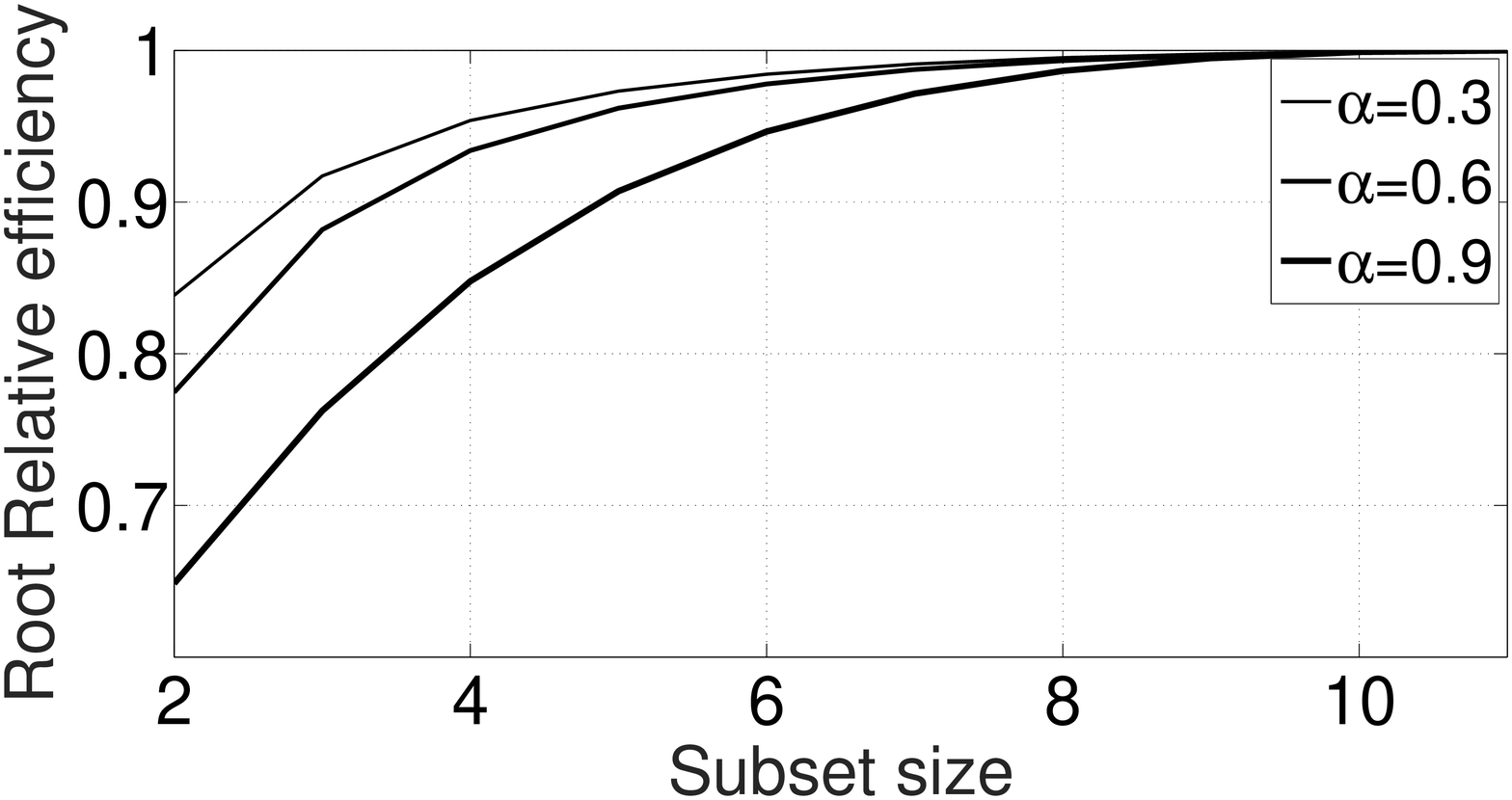}\includegraphics[width=0.53 \linewidth,height=4.6cm]{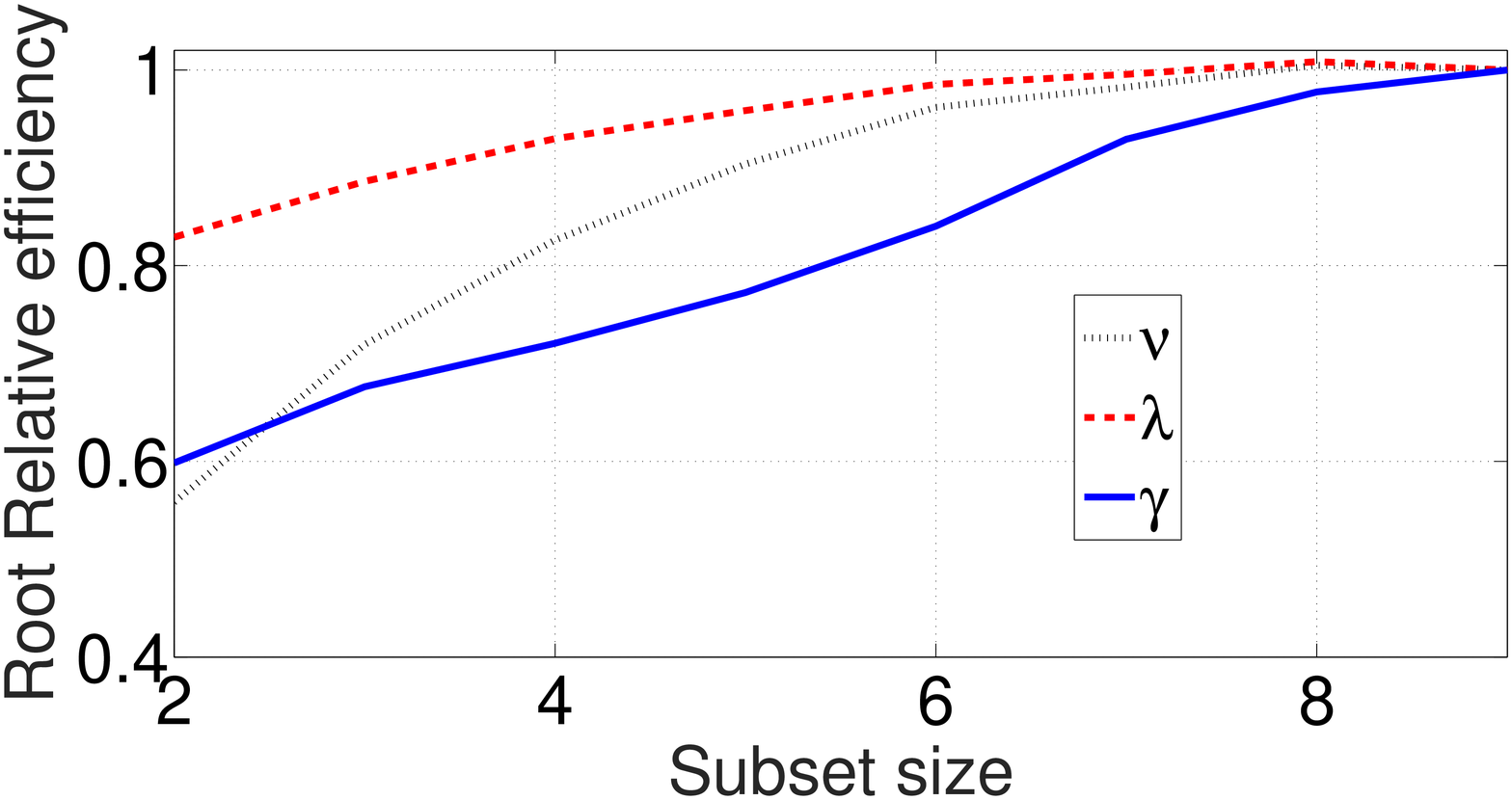}\vspace{-.18cm}}
\caption{\underline{Left:} Root relative efficiency for the simulation study for the logistic model computed over 1000 simulations of $m=50$ replicates each, with $\alpha=0.3,0.6,0.9$ (with increasingly thicker lines). The range of the absolute bias to standard deviation ratio is $(3.4\%,3.8\%)$ (median $3.8\%$), $(1.3\%,4.3\%)$ (median $4.0\%$) and $(0.0\%,8.1\%)$ (median $4.1\%$) for $\alpha=0.3,0.6,0.9$ respectively. The plot of the standard deviation ratio (not shown) is graphically undistinguishable.
\underline{Right:} Root relative efficiency of complete composite likelihood estimators for $\lambda$ (in dashed red) and for $\nu$ (in dotted black), for 100 experiments of $m=50$ replicates of the Brown--Resnick process with semi-variogram $\gamma(\mbf{h})=(\|\mbf{h}\|/\lambda)^{\nu}$ for $\lambda=0.42$ and $\nu=1.5$, observed at nine locations. Results are plotted against the subset size $q$ of composite likelihoods. The range of the absolute bias to standard deviation ratio is $(1.9\%,13.5\%)$ (median $12.2\%$) for $\alpha$ and $(9.7\%,15.7\%)$ (median $14.2\%$) for $\lambda$. The plot of the standard deviation ratio (not shown) is graphically undistinguishable. The ratio of $L^2$ distances (taken over the range from $\|\mbf{h}\|=0$ to $\|\mbf{h}\|=\sqrt{2}$) between the estimated and the true semi-variogram $\gamma(\mbf{h})$ at $Q=9$ as a function of $q$ is also shown in solid blue.}
\label{logistic_rmse}
\end{figure}

\paragraph{The logistic model.}\label{log_stud}
For $\alpha=0.3,0.6,0.9$ (strong, mild and weak dependence, respectively), we perform 1000 independent experiments. For each experiment $j=1,\ldots,1000$, we simulate $m=50$ independent replicates (a realistic number for most environmental applications) from a $Q=11$-dimensional logistic distribution and obtain the maximum complete composite likelihood estimator of order $q$, $\hat{\alpha}_{j,q}$, from \eqref{likelihoods} with subsets of cardinality $q=2,\ldots,Q$. Following \cite{hu13}, the root mean squared error $\text{rmse}_q=\sqrt{\text{b}_q^2+\text{sd}^2_q}$ is computed, where the bias is $\text{b}_q=\bar{\hat{\alpha}}_q-\alpha$, ($\bar{\hat{\alpha}}_q=\sum_{j=1}^{1000} \hat{\alpha}_{j,q}/1000$), and the standard deviation is $\text{sd}_q=\sqrt{\sum_{j=1}^{1000}(\hat{\alpha}_{j,q}-\bar{\hat{\alpha}}_q)^2/999}$. The root relative efficiencies defined as $\text{rre}_q=\text{rmse}_Q/\text{rmse}_q$ are shown in Figure \ref{logistic_rmse} (left), and the range (and the sample median) of absolute bias to standard deviation ratio is reported in the caption. Overall, complete composite likelihood estimators improve as $q$ increases. Moreover, as the dependence decreases (i.e., $\alpha$ increases), the gain of high-order composite likelihood estimators with respect to low-order counterparts is more evident. Overall, the root mean squared error of the full likelihood estimator is between $16\%$ and $35\%$ (respectively, between $8\%$ and $23\%$) smaller than that of the pairwise (respectively triplewise) complete composite likelihood estimator.

\vspace{-.4cm}

\paragraph{The Reich--Shaby model.}\label{alog_stud}
The logistic model (\ref{logistic}) considered previously applies to multivariate extremes, but has no notion of spatial dependence. As discussed in Section~\ref{ms_proc}, the Reich--Shaby model is a spatial extension of the logistic model, whose exponent measure and derivatives are relatively convenient to compute (see Supplementary Material). 

To be consistent with the logistic model, 1000 experiments are performed, each with $m=50$ independent replicates. For each experiment, $Q=11$ stations are uniformly generated on $[0,1]^2$, and data are simulated at those (fixed) locations with knots $\mbf{v}_1,\ldots,\mbf{v}_{36}$ on a regular grid $[0,0.2,\ldots,1]^2$. Different levels of noise, with $\alpha=0.3,0.6,0.9$ (little to very noisy), and different dependence ranges, with $\tau=0.1,0.2,0.4$ (short- to long-range), are considered. Holding knots fixed, we estimate the two parameters $\alpha$ and $\tau$ using maximum complete composite likelihoods estimators of order $q=2,\ldots,Q$, defined in \eqref{likelihoods}. The root relative efficiency $\text{rre}_q$ is reported in Table~\ref{alog:results_alpha}. Estimates of $\alpha$ show a strictly monotonic improvement with higher-order composite likelihoods, as noticed for the logistic case, while for $\tau$ the efficiency is not strictly monotonic for high $q$s. Overall, the root mean squared error of the full likelihood estimator is $20\%$ to $45\%$ (respectively, $14\%$ to $31\%$) smaller than that of the pairwise (respectively triplewise) complete composite likelihood estimator for $\hat{\alpha}$ and $6\%$ to $71\%$ (respectively, $7\%$ to $66\%$) for~$\hat{\tau}$. 

\begin{table}[h!]
\caption{{\scriptsize Root relative efficiency $(\%)$ for $\hat \alpha$ and $\hat \tau$ computed over 1000 simulations of $m=50$ replicates each, for different values of $\alpha$ and $\tau$ in the Reich--Shaby model. The range of the absolute bias to standard deviation ratio is $(0.2\%,27.4\%)$ (median $6.0\%$) for $\alpha$ and $(0.9\%,17.8\%)$ (median $7.9\%$) for $\tau$.}\vspace{-.2cm}}\label{alog:results_alpha}
\scriptsize
\centering
\centerline{$\hat \alpha / \hat \tau$}
\vspace{5pt}
\begin{tabular}{|c|c||c|c|c|c|c|c|c|c|c|c|}
\hline
$\tau$ & $\alpha$   & $q=2$ & $q=3$ & $q=4$ & $q=5$ & $q=6$ & $q=7$ & $q=8$ & $q=9$ & $q=10$ & $q=11$\\ \cline{1-12} 

   & 0.3 &  76/65 &  82/77 &  87/82 &  90/87 &  93/91 &  95/95 &  97/97 &  99/99 & 100/100 & 100/100  \\ \cline{2-12}
0.1 & 0.6 &  77/55 &  83/71 &  87/78 &  90/84 &  92/89 &  94/95 &  96/96 &  98/97 &  99/99 & 100/100  \\ \cline{2-12}
   & 0.9 &  79/74 &  83/86 &  84/82 &  85/91 &  81/89 &  90/93 &  90/89 &  96/100 &  97/97 & 100/100  \\ \cline{1-12}
    & 0.3 &  55/29 &  69/44 &  79/62 &  85/75 &  90/84 &  94/91 &  96/95 &  98/98 & 100/99 & 100/100 \\ \cline{2-12}
0.2 & 0.6 &  75/54 &  84/64 &  89/74 &  92/82 &  95/88 &  97/93 &  98/96 &  99/97 & 100/98 & 100/100  \\ \cline{2-12}
    & 0.9 &  80/72 &  84/80 &  87/83 &  89/79 &  92/87 &  94/94 &  95/98 &  96/103 &  98/105 & 100/100 \\ \cline{1-12}
    & 0.3 & 72/56 &  81/63 &  87/70 &  92/76 &  95/82 &  97/87 &  98/92 &  99/95 & 100/97 & 100/100  \\ \cline{2-12}
0.4 & 0.6 & 77/74 &  86/86 &  91/91 &  95/94 &  97/97 &  99/98 & 100/100 & 100/100 & 100/101 & 100/100  \\ \cline{2-12}
    & 0.9 & 59/94 &  83/93 &  72/97 &  74/100 &  76/100 &  71/100 &  99/106 & 100/103 & 100/105 & 100/100   \\ \cline{2-12}
\hline
\end{tabular}
\end{table}

Besides the increase in root relative efficiency, we found evidence that the use of high-order likelihoods also results in a decrease in correlation between $\hat{\alpha}$ and $\hat{\tau}$, especially when this correlation is nonnegligible (see Supplementary Material). 

\vspace{-.4cm}

\paragraph{The Brown--Resnick model.}\label{br_sim_c}

After an experiment on a relatively simple model for spatial extremes, we conduct a simulation study in a more realistic and more computationally intensive context: 100 experiments are performed, each consisting of $m=50$ replicates of a Brown--Resnick process with semi-variogram $\gamma(\mbf{h})=(\|\mbf{h}\|/\lambda)^{\nu}$ and $\lambda=0.42$, $\nu=1.5$ (a case considered by \citealp{hu13}, and illustrated in Figure \ref{real_plot}(g)), on $Q=9$ stations uniformly generated in the unit square. The dimensionality and number of experiments are reduced from the two previous settings since evaluating partial derivatives of the exponent measure for the Brown--Resnick model requires expensive computations of high-dimensional normal cumulative distribution functions (see \eqref{BRmodel} and the Supplementary Material), which rely on quasi-Monte Carlo approaches \citep{ge92,ge02,ge09}. The range $\lambda>0$ and smoothness $\nu\in(0,2]$  parameters are estimated using maximum complete composite likelihoods estimators of order $q=2,\ldots,Q$. Since the complete simulation study in this case requires approximately two weeks, it was not possible to repeat it for other parameter values. As we can see from the results in Figure \ref{logistic_rmse} (right), both parameters show an increase in efficiency with subset size $q$. This is especially striking for $\nu$: the pairwise (respectively, triplewise) likelihood estimator is approximately $44\%$ (respectively $28\%$) less efficient than the full likelihood estimator. By comparison, the loss in efficiency for the range $\lambda$ is about $17\%$ (respectively $13\%$) for pairwise (respectively triplewise) likelihood estimators. High values of $q$ did not result in a noticeable change in the estimated (negative) correlation between $\hat{\lambda}$ and $\hat{\nu}$ (results not shown). Besides the interest in parameter estimation, we also quantify how close the estimated spatial structure is to the true one. In Figure \ref{logistic_rmse} (right), we report $\|\gamma-\hat{\gamma}_Q\|_2/\|\gamma-\hat{\gamma}_q\|_2$, where $\|\cdot\|_2$ denotes the $L^2$ norm taken over the range from $\|\mbf{h}\|=0$ to $\|\mbf{h}\|=\sqrt{2}$, and $\hat{\gamma}_q$ is the variogram estimated with a composite likelihood of order $q$. It is apparent that the use of high-order likelihoods results in a better agreement with the true spatial dependence, and that the increase in efficiency is more striking compared to parameter estimates. Overall, the patterns for $\nu$ and $\lambda$ seem coherent with what was obtained for the logistic and the Reich--Shaby models, so we believe these results can be extrapolated by analogy to Brown--Resnick processes with other parameter combinations.

\vspace{-.18cm}

\subsection{Truncated Composite Likelihoods}\label{lik_appr}

In Section~\ref{complete_complik}, we showed that the use of high-order complete composite likelihoods, or full likelihoods when possible, results in a better estimation performance. However, to further improve the estimation and to reduce the computational burden, an option is to consider the larger class of weighted composite likelihoods and to investigate how to make the best choice of weights in \eqref{ms_density_composite}. One solution could be to choose weights minimizing the trace or the determinant of the asymptotic variance of $\hat{\bsy{\theta}}_C$ \citep{pa10,sa14}, but this is challenging to implement and computationally demanding. Furthermore, if weights are all non-zero, this approach might improve the estimation efficiency, but it does not reduce the computational time, since it still requires evaluating all the elements in $C_{\mbf{z}_q}(Q)$. An alternative solution is to select only some elements of the collection $C_{\mbf{z}_q}(Q)$ in \eqref{ms_density_composite} (i.e., choose binary weights) depending on their set distances, on the ground that close locations are generally more informative about dependence parameters than are distant, less correlated, ones. We propose to rank all elements in $C_{\mbf{z}_q}(Q)$ according to their maximum set distance (i.e., the maximum distance among all pairs in the subset), and to consider only a percentage of the ranked vector.

To investigate the efficiency of this approach, we perform $1000$ experiments with $m=50$ replicates of the Reich--Shaby model. For $t=0.1,0.2,\ldots,1$, we consider the first $[t\times |C_{\mbf{z}_q}(Q)|]$ elements of the ranked collection $C_{\mbf{z}_q}(Q)$, and we compute the root relative efficiency $\text{rre}_q$ (with respect to the full likelihood) as in Section \ref{complete_complik}. In other words, we increase the number of elements of $C_{\mbf{z}_q}(Q)$ from $10\%$ to $100\%$ and investigate how the root mean squared error is affected for different values of $q$. The $100\%$ case considers the whole set $C_{\mbf{z}_q}(Q)$ and therefore corresponds to the complete composite likelihood case of Section~\ref{complete_complik}. This simulation study requires maximizing a likelihood for every value of $t$, of $q=2,\ldots,Q-1$ with $Q=11$ (the full likelihood case for $q=Q$ has $|C_{\mbf{z}_Q}(Q)|=1$) and experiment $j=1,\ldots,1000$, for a total of about $10^5$ maximizations. This simulation study requires a week of computation on the fully dedicated cluster for each parameter choice. Table \ref{alog:results_alpha_trunc} reports the results for $\alpha=0.6$ and $\tau=0.2$, but similar patterns were observed for all other parameter configurations. Regarding the estimation of $\alpha$, the most efficient estimators are for $t=100\%$ when $q\geq 6$. For $q=2$, the best estimate is at $t=70\%$  (which confirms that the complete pairwise likelihood is not always the best solution, as observed by \citealp{sa14}), while for $q=3,4,5$ the best is at $t=50\%,70\%,90\%$. 
\begin{table}[t!]
\caption{{\scriptsize Root relative efficiency with respect to the full likelihood for $\hat \alpha$ and $\hat \tau$, computed over 1000 simulations of $m=50$ replicates each of the Reich--Shaby model with $\tau=0.2$ and $\alpha=0.6$, considering $[t\times |C_{\mbf{z}_q}(Q)|]$ elements $t=0.1,0.2,\ldots,1$. In bold, the maximum efficiency across $t$ for all composite likelihood orders. In the last two rows, the smallest $t$ (\%) for the $q$-set in order to beat the best result for $(q-1)$-set. In parentheses, the ratio ($\times100$) of the elapsed times (averaged across experiments) between these two combinations: values less than 100 mean that it is less time demanding to use an optimal $t$ for $(q-1)$-sets rather than considering $q$-sets, and vice versa. The range of the absolute bias to standard deviation ratio is $(0.0\%,20.2\%)$ (median $3.3\%$) for $\alpha$ and $(6.2\%,21.0\%)$ (median $12.7\%$) for $\tau$.}\vspace{-.3cm}}\label{alog:results_alpha_trunc}
\scriptsize
\centering
\centerline{$\hat \alpha / \hat \tau$}
\vspace{1pt}
\begin{tabular}{|c||c|c|c|c|c|c|c|c|c|}
\hline
\backslashbox{$t$ (\%)}{$q$}    & 2 & 3 & 4 & 5 & 6 & 7 & 8 & 9 & 10 \\ \cline{1-10} 
10   &  52/15 &  79/48 &  84/66 &  87/72 &  89/76 &  92/82 &  94/87 &  95/92 &  96/89  \\ \cline{1-10}  
20   &  63/22 &  84/64 &  87/72 &  89/75 &  92/81 &  94/87 &  96/90 &  96/92 &  98/94  \\ \cline{1-10}  
30   &  70/36 &  85/69 &  87/74 &  91/78 &  93/86 &  95/88 &  96/92 &  98/93 &  98/95  \\ \cline{1-10}  
40   &  74/46 &  86/\textbf{70} &  89/74 &  91/80 &  94/86 &  95/89 &  97/93 &  98/95 &  98/97  \\ \cline{1-10}  
50   &  78/54 &  \textbf{86}/69 &  89/74 &  92/81 &  94/86 &  96/90 &  97/93 &  98/96 & 100/98  \\ \cline{1-10}  
60   &  79/\textbf{61} &  86/68 &  89/\textbf{75} &  92/81 &  94/87 &  96/90 &  98/94 &  99/97 & 100/98  \\ \cline{1-10}  
70   &  \textbf{79}/61 &  85/67 &  \textbf{89}/74 &  92/81 &  95/87 &  97/91 &  98/95 &  99/97 & 100/97  \\ \cline{1-10}  
80   &  78/58 &  85/66 &  89/74 &  92/81 &  95/87 &  97/92 &  98/95 &  99/98 & 100/97  \\ \cline{1-10}  
90   &  77/56 &  84/65 &  89/74 &  \textbf{93}/\textbf{82} &  95/88 &  97/92 &  98/\textbf{96} &  99/\textbf{98} & 100/\textbf{98}  \\ \cline{1-10}  
100   & 75/53 &  84/64 &  89/74 &  93/82 &  \textbf{95}/\textbf{88} &  \textbf{97}/\textbf{92} &  \textbf{98}/95 &  \textbf{99}/97 & \textbf{100}/98 \\ \cline{1-10}  
\hline\hline
$\hat{\alpha}$   &   &  10(99) &  20(61) &  20(116) &  20(202) &  30(199) &  40(187) &  60(148) &  80(136)    \\ \cline{1-10}  
\hline
$\hat{\tau}$   &  &  20(50) &  20(49) &  20(100) &  20(202) &  20(298) &  40(187) &  50(157) &  40(266)   \\ \cline{1-10}   
\end{tabular}
\end{table}
Therefore, truncating $C_{\mbf{z}_q}(Q)$ slightly improves the estimation when $q$ is small, but for higher orders, complete composite likelihoods give better results. In the last two rows, we also report the smallest truncation factor of $t$ for the $q$-set that corresponds to the best efficiency result with the best $(q-1)$-set and in parentheses the ratio ($\times100$) of the elapsed times (averaged across experiments) between these two combinations. For $q=3,4$, an optimal choice of the truncation for $q-1$ allows for faster likelihood evaluation than considering $q$-sets. This is not the case when $q\geq 5$, but the efficiency gain at these cardinalities is considerably smaller than for small $q$s. For the estimation of $\tau$, the truncation improves estimation when subsets are of size $q=2$--$5,8,9$, although the optimal estimators for a given $q$ are never better than the ones for $q+1$ at $t=100\%$. From the last row, we see that it is more efficient to choose an optimal $t$ for a given $q-1$ than evaluating the composite likelihood for $q$ if $q=3,4$, while the opposite is true for $q\geq 5$. We therefore conclude that, for low-order composite likelihoods, truncation can improve the estimation precision while also decreasing the computation, especially for~$\tau$.

A similar study was carried out for the Brown--Resnick process, with results shown in tables similar to Table~\ref{alog:results_alpha_trunc} and reported in the Supplementary Material. Similarly to the Reich--Shaby case, the best relative efficiency is achieved once a truncation is performed, and there is a diminished improvement in performance with a truncation of high-order composite likelihoods for both $\lambda$ and $\nu$. Similarly, for low-order composite likelihoods a truncation at $q-1$ is computationally more convenient than considering likelihoods of order $q$.

\section{Computational Cost of Composite Likelihoods}\label{cost}
\subsection{Memory and CPU requirements}

\begin{figure}[b!]
\centerline{\includegraphics[width=11cm,keepaspectratio]{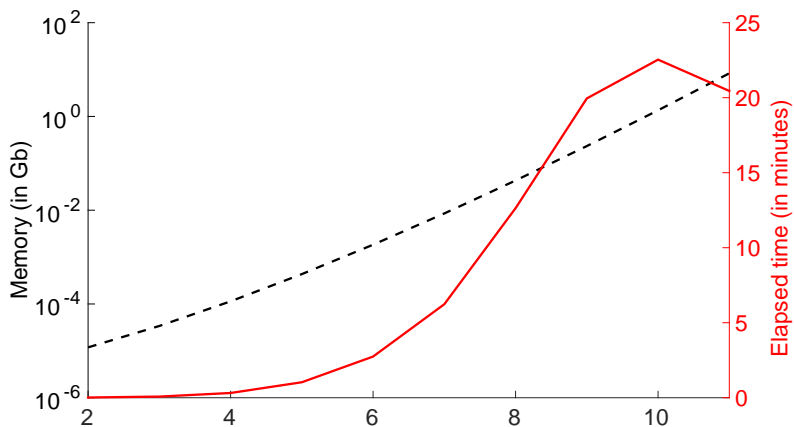}\vspace{-.2cm}}
\caption{Average memory required (on log scale and in dashed black) and elapsed time (in solid red) across simulations per individual likelihood evaluation. Simulation studies for 1000 experiments, each one with $m=50$ replicates, for the logistic model with $\alpha=0.6$. }
\label{log_sim}
\end{figure}

As demonstrated in Section~\ref{sim_stud}, high-order composite likelihood estimators generally perform better than pairwise or triplewise approaches, in terms of root mean squared error. This improved performance, however, has a computational cost, which increases rapidly with the composite likelihood order $q$.

To illustrate this fact, Figure \ref{log_sim} displays the computational requirements of complete composite likelihood estimators as a function of the order $q$, for the simulation study described in Section~\ref{complete_complik} for the logistic model with $\alpha=0.6$. The figure shows the memory required to store the cell array containing all partitions, $\mbf{P}_q$, and the elapsed time (averaged across experiments) for different values of $q$. This shows that the required memory, which does not depend on the model, is the main limitation for the use of high-order composite likelihoods; in comparison, for the logistic model, the computational complexity is not as problematic (but it could be for other models). Composite likelihoods of order $q$ require storing $\mbf{P}_q$ containing $B_q=O(q^q)$ elements and cannot be larger than $B_{11}=678,570$ with our facilities (since a copy of $\mbf{P}_q$ had to be created for each of the 780 CPUs used for parallel computing), although if one uses shared memory of if the goal is to analyze a single data set, $q$ could be 12 or 13.  

A possible solution could be to use the divide-and-conquer strategy by splitting the computation for all partitions according to some criterion, the most natural being the number of sets within the partition (so-called Stirling partitions, see \citealp{gr88}). This would mitigate the storage problem, but the dynamical computation of the subsets has proven to be, in our experience, very inefficient. This suggests that even with an efficient coding on a powerful computer, the full likelihood on a single data set may be computed in dimensions $Q=12$ or $Q=13$, and similarly, that composite likelihoods are limited to order $q=13$. Given the very steep increase in memory usage with $q$, these estimated upper bounds should not change noticeably in coming years.

Figure \ref{log_sim} also reveals an interesting fact concerning complete composite likelihoods: the computational time does not always increase monotonically with subset size. The full likelihood approach is less demanding than 9-sets or 10-sets because an evaluation with 9- or 10-sets requires the computation of ${11\choose 9}\times B_9$ or ${11\choose 10}\times B_{10}$ terms in (\ref{ms_density_composite}), respectively, which is more computationally demanding than summing over $B_{11}$ terms once.
\vspace{-.2cm}

\subsection{Projections for higher dimensions}

While the results discussed in Section \ref{sim_stud} and the computational limitations in the previous subsection are important to understand the benefits of high-order likelihood inference, they do not provide insights on how to perform an analysis with a large data set. Although an analysis with $q$ larger than 13 is not feasible with current computers, an important question concerns computational requirements for composite likelihood evaluation with relatively small $q$ but large $Q$. 
Table \ref{log:appl} shows the estimated elapsed time per likelihood evaluation on a single CPU in the setting of Section \ref{log_stud} in the logistic case (a similar table for the Reich--Shaby model can be found in the Supplementary Material), with $m=50$ and $\alpha=0.6$, for different values of $q$ and $Q$. The elapsed time $e_{q,Q}$ is computed directly if it is less than 25 seconds. For larger computational times, we estimate it as $e_{q,\tilde{Q}}\times\frac{|C_{\mbf{z}_q}(Q)|}{|C_{\mbf{z}_q}(\tilde{Q})|}$, where $\tilde{Q}$ is the largest $Q$ for which the computation requires less than 25 seconds. Since increasing $Q$ will increase the size of $C_{\mbf{z}_q}(Q)$, but not the computational cost per element, we believe this simple extrapolation provides a realistic indication of the projected time for higher dimensions. Whenever the projected time is more than one day, it is clearly not feasible to perform a likelihood maximization, and we therefore also report the truncation proportion $t$ (recall Section~\ref{lik_appr}) needed to reduce the projected time to one day.

\begin{table}[b!]
\caption{{\scriptsize Estimated elapsed time per likelihood evaluation on a single CPU, for different values of $q$ and $Q$ for $m=50$ in the logistic case (s=seconds, m=minutes, h=hours, d=days). When the required time is more than 1 day, the required truncation (in $\%$) to decrease the computational time to this threshold are in parenthesis. Estimated times over 1 month are indicated as~$>30$d, and below 1 second as~$<1s$.}}\label{log:appl}
\vspace{-.2cm}
\scriptsize\centering
\begin{tabular}{|c||c|c|c|c|c|c|c|c|c|}
\hline
\backslashbox{$Q$}{$q$}   & 2 & 3 & 4 & 5 & 6 & 7 & 8 \\ \cline{1-8} 
11 & $<$1s & $<$1s &1s &2s &4s &10s &19s \\ \cline{1-8}
15 & $<$1s & $<$1s &3s &13s &44s &3m &12m \\ \cline{1-8}
20 & $<$1s &1s &8s &1m &6m &40m &4h \\ \cline{1-8}
50 & $<$1s &14s &6m &2h &2d(62) & $>$30d(3) & $>$30d(0.14) \\ \cline{1-8}
100 & 2s &2m &2h &4d(27) & $>$30d(0.83) & $>$30d(0.02) & $>$30d(0) \\ \cline{1-8}
500 & 54s &4h & $>$30d(2) & $>$30d(0.01) & $>$30d(0) & $>$30d(0) & $>$30d(0) \\ \cline{1-8}
1,000 & 4m &1d(71) & $>$30d(0.13) & $>$30d(0) & $>$30d(0) & $>$30d(0) & $>$30d(0) \\ \cline{1-8}
5,000 & 2h & $>$30d(0.56) & $>$30d(0) & $>$30d(0) & $>$30d(0) & $>$30d(0) & $>$30d(0) \\ \cline{1-8}
10,000 & 6h & $>$30d(0.07) & $>$30d(0) & $>$30d(0) & $>$30d(0) & $>$30d(0) & $>$30d(0) \\ \cline{1-8}
100,000 & 25d(4) & $>$30d(0) & $>$30d(0) & $>$30d(0) & $>$30d(0) & $>$30d(0) & $>$30d(0) \\ \cline{1-8}
\hline
\end{tabular}
\vspace{-.5cm}
\end{table}

It is clear from the results that when $Q$ is large, the complete composite likelihood evaluation becomes problematic, even for small orders $q$. By contrast, when $Q=11,15,20$, it is possible to evaluate composite likelihoods with $q=8$ in a relatively short time without truncation. As the dimension $Q$ increases, a truncation may be necessary to reduce the computational time (especially for high $q$s), and for very large dimensions ($Q=10,000,\,100,000$), pairwise likelihood seems to be the only viable solution. This table was obtained with a single 2.4GHz processor, and the results should not improve significantly with multiple processors. This is because $|C_{\mbf{z}_q}(Q)|$ becomes very large for even small values of $Q$, therefore requiring a very large number of fast independent operations. Under this setting, parallel computing will likely perform worse than a single CPU.
\vspace{-.1cm}

\subsection{Recommendations to practitioners}
\vspace{-.2cm}

We now summarize some findings, which may be helpful to practitioners, who have to find the best compromise between statistical and computational efficiency:
\vspace{-.25cm}
\begin{itemize}
\item Truncation decreases the computational time and also improves the statistical efficiency for low-order composite likelihoods.\vspace{-.2cm}

\item Computational time may also be decreased by providing good starting values to maximize more efficient higher-order composite likelihoods, for example by using low-order composite likelihood estimates as starting values.\vspace{-.2cm}

\item For small dimensional data sets ($Q=11,15,20$), a high-order composite likelihood is possible to compute and improves the inference, although there is a diminished return as $q$ increase and the memory usage should be always monitored. For high dimensional data sets ($Q>20$), composite likelihoods can be computed only for relatively small $q$s (unless a hard truncation is applied).\vspace{-.2cm}

\item When the dimensionality is too high for a complete composite likelihood evaluation, as it is likely the case in typical environmental applications, a truncation is absolutely needed. The choice of the truncation proportion will likely be dictated from computational convenience rather than parameter estimates optimality, but truncated high-order composite likelihoods also result in better performance than a complete low-order composite likelihood.  
\end{itemize}

\section{Discussion}\label{concl}
In this work, we have tackled the challenging problem of inference for max-stable processes from a computational perspective and have explored the limits of likelihood-based inference. We have shown that high-performance computing can be a powerful tool to perform efficient inference for multivariate or spatial extremes, but even with a large memory and an efficient use of computational resources, full likelihood inference seems to be limited to dimension $Q=12$ or $Q=13$ with current technologies; for higher dimensions, composite likelihoods with suitably selected components are an efficient alternative, although, for the same reasons, they are limited to order $q=12$ or $q=13$. With extensive simulation studies based on three increasingly complex classes of multivariate or spatial extremes, we have quantified the loss of efficiency of widely-used pairwise and triplewise likelihood approaches and found that high-order composite likelihoods can lead to substantial improvements in root mean squared error. Furthermore, we have shown that truncation of composite likelihoods not only reduces the computational burden, but in some cases also leads to better estimation of the parameters, especially for low-order composite likelihoods, confirming results obtained in previous studies. Finally, we have also given guidance on the choice of subsets to include into truncated composite likelihoods, advocating a simpler approach based on the maximum subset distance. This choice is, however, not optimal and other possibilities could be to choose the subsets based on the average, or minimal, subset distance.

We do not expect that computers with contemporary architectures will render the ``full likelihood problem'' easy to solve for large dimensions in the foreseeable future: an increase in available processors would only linearly decrease the computational time, and memory storage is not foreseen to increase enough to allow the storage of, say, data structures with $B_{30}= O(10^{23})$ elements. We therefore conclude that a direct full likelihood approach is not feasible unless important methodological advances are made to deal with this problem, and truncated composite likelihoods with subsets of size $q=4$--$6$ provide an efficient and relatively cheap solution in the settings we have examined. For very large data sets, however, the only solution might be to use pairwise likelihoods.

Given the computationally demanding nature of these approaches, it is natural to wonder whether other likelihoods can be derived from alternative representations of extreme events. We have assumed that max-stable data (typically pointwise maxima of random processes) are the only information available for inference. However, if additional information, such as the times of occurrence of maxima or the original processes, is incorporated, considerably more efficient strategies based on point processes can be devised; see \citet{st05,wa14,thi15}. Alternatively, recent advances in the simulation of max-stable processes \citep{Dieker.Mikosch:2014,Dombry.etal:2015} suggest that simulation-based likelihood inference might be an alternative \citep{ko14}.

In this work, we focused on three max-stable parametric families, but the considerations on computational feasibility are the same for every process whose cumulative distribution function has the form \eqref{distrMS}. Furthermore, we believe that similar efficiency results should be expected for composite likelihoods applied to other max-stable models, such as the \cite{Schlather:2002} or the extremal $t$ model \citep{Nikoloulopoulos.etal:2009,op13}, under similar scenarios. It would also be worth investigating the case of multivariate max-stable processes \citep{Oesting.etal:2013,ge14}, or asymptotically independent models \citep{Wadsworth.Tawn:2012}, where the rate of decay towards independence has to be estimated.

\section*{Supplementary Materials}

\noindent {\bf Supplements:} Correlation between $\hat \alpha$ and $\hat \tau$ in the Reich--Shaby model; Performance of truncated composite likelihood
estimators for the Brown--Resnick model; Computational time projections for the Reich--Shaby model; Partition-based likelihood approximation; Closed form expression of the partial derivatives for the three models (SupplementsCHG.pdf)

\noindent {\bf Code:} Simulation code (CodeCHG.zip)

\baselineskip 21.3 pt
\bibliographystyle{asa}
\bibliography{citations}

\begin{thebibliography}{48}
\newcommand{\enquote}[1]{``#1''}
\expandafter\ifx\csname natexlab\endcsname\relax\def\natexlab#1{#1}\fi

\bibitem[{Ballani and Schlather(2011)}]{Ballani.Schlather:2011}
Ballani, F. and Schlather, M. (2011), \enquote{{A Construction Principle for
  Multivariate Extreme Value Distributions},} \textit{{Biometrika}}, 98,
  633--645.

\bibitem[{Bevilacqua et~al.(2012)Bevilacqua, Gaetan, Mateu, and Porcu}]{be12}
Bevilacqua, M., Gaetan, C., Mateu, J., and Porcu, E. (2012),
  \enquote{Estimating Space and Space-Time Covariance Functions for Large Data
  Sets: A Weighted Composite Likelihood Approach,} \textit{Journal of the
  American Statistical Association}, 107, 268--280.

\bibitem[{Bienven\"ue and Robert(2014)}]{bi14}
Bienven\"ue, A. and Robert, C.~Y. (2014), \enquote{Likelihood Based Inference
  for High-dimensional Extreme Value Distributions,} {arXiv:1403.0065v3}.

\bibitem[{Brown and Resnick(1977)}]{br77}
Brown, B.~M. and Resnick, S.~I. (1977), \enquote{Extreme Values of Independent
  Stochastic Processes,} \textit{Journal of Applied Probability}, 14, 732--739.

\bibitem[{Coles(2001)}]{co01}
Coles, S. (2001), \textit{An Introduction to Statistical Modeling of Extreme
  Values}, New York: Springer.

\bibitem[{Cooley et~al.(2010)Cooley, Davis, and Naveau}]{Cooley.etal:2010}
Cooley, D., Davis, R.~A., and Naveau, P. (2010), \enquote{{The Pairwise Beta
  Distribution: A Flexible Parametric Multivariate Model for Extremes},}
  \textit{{Journal of Multivariate Analysis}}, 101, 2103--2117.

\bibitem[{Cox and Reid(2004)}]{co04}
Cox, D.~R. and Reid, N. (2004), \enquote{A Note on Pseudolikelihood Constructed
  from Marginal Densities,} \textit{Biometrika}, 91, 729--737.

\bibitem[{Davis and Yau(2011)}]{da11}
Davis, R. and Yau, C. (2011), \enquote{Comments on Pairwise Likelihood in Time
  Series Models,} \textit{Statistica Sinica}, 21, 255--278.

\bibitem[{Davison et~al.(2012)Davison, Padoan, and Ribatet}]{da12b}
Davison, A., Padoan, S.~A., and Ribatet, M. (2012), \enquote{Statistical
  Modeling of Spatial Extremes,} \textit{Statistical Science}, 27, 161--186.

\bibitem[{Davison and Huser(2015)}]{Davison.Huser:2015}
Davison, A.~C. and Huser, R. (2015), \enquote{{Statistics of Extremes},}
  \textit{{Annual Review of Statistics and its Application.}}, 2, 203--235.

\bibitem[{Davison et~al.(2013)Davison, Huser, and Thibaud}]{Davison.etal:2013}
Davison, A.~C., Huser, R., and Thibaud, E. (2013), \enquote{{Geostatistics of
  Dependent and Asymptotically Independent Extremes},} \textit{{Mathematical
  Geosciences}}, 45, 511--529.

\bibitem[{Dieker and Mikosch(2015)}]{Dieker.Mikosch:2014}
Dieker, A. and Mikosch, T. (2015), \enquote{{Exact Simulation of Brown--Resnick
  Random Fields at a Finite Number of Locations},} \textit{{Extremes}}, 18,
  301--314.

\bibitem[{Dombry et~al.(2015)Dombry, Engelke, and Oesting}]{Dombry.etal:2015}
Dombry, C., Engelke, S., and Oesting, M. (2015), \enquote{{Exact Simulation of
  Max-Stable Random Fields},} {arXiv:1506.04430v1}.

\bibitem[{Eidsvik et~al.(2014)Eidsvik, Shaby, Reich, Matthew, and
  Niemi}]{Eidsvik.etal:2014}
Eidsvik, J., Shaby, B.~A., Reich, B.~J., Matthew, W., and Niemi, J. (2014),
  \enquote{{Estimation and Prediction in Spatial Models With Block Composite
  Likelihoods},} \textit{{Journal of Computational and Graphical Statistics}},
  23, 295--315.

\bibitem[{Genton et~al.(2011)Genton, Ma, and Sang}]{gen11}
Genton, M.~G., Ma, Y., and Sang, H. (2011), \enquote{On the Likelihood Function
  of Gaussian Max-stable Processes,} \textit{Biometrika}, 98, 481--488.

\bibitem[{Genton et~al.(2015)Genton, Padoan, and Sang}]{ge14}
Genton, M.~G., Padoan, S., and Sang, H. (2015), \enquote{Multivariate
  Max-Stable Spatial Processes,} \textit{Biometrika}, 102, 215--230.

\bibitem[{Genz(1992)}]{ge92}
Genz, A. (1992), \enquote{Numerical Computation of Multivariate Normal
  Probabilities,} \textit{Journal of Computational and Grapical Statistics}, 1,
  141--149.

\bibitem[{Genz and Bretz(2002)}]{ge02}
Genz, A. and Bretz, F. (2002), \enquote{Methods for the Computation of
  Multivariate t-probabilities,} \textit{Journal of Computational and Grapical
  Statistics}, 11, 950--971.

\bibitem[{Genz and Bretz(2009)}]{ge09}
--- (2009), \textit{Computation of Multivariate Normal and t Probabilities},
  Berlin: Springer.

\bibitem[{Graham et~al.(1988)Graham, Knuth, and Patashnik}]{gr88}
Graham, R., Knuth, D.~E., and Patashnik, O. (1988), \textit{Concrete
  Mathematics}, Reading MA: Addison-Wesley.

\bibitem[{Gumbel(1960)}]{gu60}
Gumbel, E.~J. (1960), \enquote{Bivariate Exponential Distributions,}
  \textit{Journal of the American Statistical Association}, 55, 698--707.

\bibitem[{Hjort and Varin(2008)}]{hj08}
Hjort, N. and Varin, C. (2008), \enquote{ML, PL, QL in Markov Chain Models,}
  \textit{Scandinavian Journal of Statistics}, 35, 64--82.

\bibitem[{Huser and Davison(2013)}]{hu13}
Huser, R. and Davison, A.~C. (2013), \enquote{Composite Likelihood Estimation
  for the Brown\mbox{-}Resnick Process,} \textit{Biometrika}, 100, 511--518.

\bibitem[{Huser et~al.(2015)Huser, Davison, and Genton}]{Huser.etal:2015}
Huser, R., Davison, A.~C., and Genton, M.~G. (2015), \enquote{{Likelihood
  Estimators for Multivariate Extremes},} {arXiv:1411.3448v2}.

\bibitem[{Joe(1997)}]{Joe:1997}
Joe, H. (1997), \textit{{Multivariate Models and Dependence Concepts}}, London:
  {Chapman \& Hall}.

\bibitem[{Kabluchko et~al.(2009)Kabluchko, Schlather, and Haan}]{ka09}
Kabluchko, Z., Schlather, M., and Haan, L. (2009), \enquote{Stationary
  Max-Stable Fields Associated to Negative Definite Functions,} \textit{The
  Annals of Probability}, 37, 2042--2065.

\bibitem[{Koch(2014)}]{ko14}
Koch, E. (2014), \enquote{{Estimation of Max-stable Processes by Simulated
  Maximum Likelihood},} Unpublished.

\bibitem[{Lindsay(1988)}]{li88}
Lindsay, B. (1988), \enquote{Composite Likelihood Methods,}
  \textit{Contemporary Mathematics}, 80, 220--239.

\bibitem[{Nikoloulopoulos et~al.(2009)Nikoloulopoulos, Joe, and
  Li}]{Nikoloulopoulos.etal:2009}
Nikoloulopoulos, A.~K., Joe, H., and Li, H. (2009), \enquote{{Extreme Value
  Properties of Multivariate $t$ Copulas},} \textit{{Extremes}}, 12, 129--148.

\bibitem[{Nychka et~al.(2015)Nychka, Bandyopadhyay, Hammerling, Lindgren, and
  Sain}]{Nychkaetal2014}
Nychka, D., Bandyopadhyay, S., Hammerling, D., Lindgren, F., and Sain, S.
  (2015), \enquote{A Multiresolution Gaussian Process Model for the Analysis of
  Large Spatial Datasets,} \textit{Journal of Computational and Graphical
  Statistics}, 24, 579--599.

\bibitem[{Oesting et~al.(2015)Oesting, Schlather, and
  Friederichs}]{Oesting.etal:2013}
Oesting, M., Schlather, M., and Friederichs, P. (2015), \enquote{{Statistical
  Post-Processing of Forecasts for Extremes Using Bivariate Brown-Resnick
  Processes with an Application to Wind Gusts},} {arXiv:1312.4584v2}.

\bibitem[{Opitz(2013)}]{op13}
Opitz, T. (2013), \enquote{Extremal t Processes: Elliptical Domain of
  Attraction and a spectral representation,} \textit{Journal of Multivariate
  Analysis}, 122, 409--413.

\bibitem[{Padoan et~al.(2010)Padoan, Ribatet, and Sisson}]{pa10}
Padoan, S.~A., Ribatet, M., and Sisson, S.~A. (2010), \enquote{Likelihood-based
  Inference for Max-stable Processes,} \textit{Journal of the American
  Statistical Association}, 105, 263--277.

\bibitem[{Reich and Shaby(2012)}]{re12}
Reich, B.~J. and Shaby, B.~A. (2012), \enquote{A Hierarchical Max-stable
  Spatial Model for Extreme Precipitation,} \textit{The Annals of Applied
  Statistics}, 6, 1430--1451.

\bibitem[{Sang and Genton(2014)}]{sa14}
Sang, H. and Genton, M. (2014), \enquote{Tapered Composite Likelihood for
  Spatial Max-stable Models,} \textit{Spatial Statistics}, 8, 86--103.

\bibitem[{Schlather(2002)}]{Schlather:2002}
Schlather, M. (2002), \enquote{{Models for Stationary Max-Stable Random
  Fields},} \textit{{Extremes}}, 5, 33--44.

\bibitem[{Segers(2012)}]{Segers:2012}
Segers, J. (2012), \enquote{{Max-stable Models for Multivariate Extremes},}
  \textit{{REVSTAT}}, 10, 61--82.

\bibitem[{Smith(1990)}]{Smith:1990}
Smith, R.~L. (1990), \enquote{{Max-stable Processes and Spatial Extremes},}
  Unpublished.

\bibitem[{Stephenson and Tawn(2005)}]{st05}
Stephenson, A. and Tawn, J.~A. (2005), \enquote{Exploiting Occurrence Times in
  Likelihood Inference for Componentwise Maxima,} \textit{Biometrika}, 1,
  213--227.

\bibitem[{Stephenson(2009)}]{st09}
Stephenson, A.~G. (2009), \enquote{High-Dimensional Parametric Modelling of
  Multivatiate Extreme Events,} \textit{Australian \& New Zealand Journal of
  Statistics}, 51, 77--88.

\bibitem[{Sun et~al.(2012)Sun, Li, and Genton}]{Sunetal2012}
Sun, Y., Li, B., and Genton, M.~G. (2012), \textit{Geostatistics for Large
  Datasets}, vol. 207, Springer, {Space-Time Processes and Challenges Related
  to Environmental Problems, E. Porcu, J. M. Montero, M. Schlather (eds),
  Chapter 3, 55--77.}

\bibitem[{Sun and Stein(2015)}]{SunStein2014}
Sun, Y. and Stein, M.~L. (2015), \enquote{Statistically and Computationally
  Efficient Estimating Equations for Large Spatial Datasets,} \textit{Journal
  of Computational and Graphical Statistics}, {in press}.

\bibitem[{Tawn(1988)}]{Tawn:1988b}
Tawn, J.~A. (1988), \enquote{{Bivariate Extreme Value Theory: Models and
  Estimation},} \textit{{Biometrika}}, 75, 397--415.

\bibitem[{Tawn(1990)}]{ta90}
--- (1990), \enquote{Modelling Multivariate Extreme Value Distributions,}
  \textit{Biometrika}, 77, 245--253.

\bibitem[{Thibaud et~al.(2015)Thibaud, Aalto, Cooley, Davison, and
  Heikkinen}]{thi15}
Thibaud, E., Aalto, J., Cooley, D.~S., Davison, A.~C., and Heikkinen, J.
  (2015), \enquote{Bayesian Inference for the Brown-Resnick Process, with an
  Application to Extreme Low Temperatures,} ArXiv:1506.07836.

\bibitem[{Varin et~al.(2011)Varin, Reid, and Firth}]{va11}
Varin, C., Reid, N., and Firth, D. (2011), \enquote{An Overview of Composite
  Likelihood Methods,} \textit{Statistica Sinica}, 21, 5--42.

\bibitem[{Wadsworth and Tawn(2012)}]{Wadsworth.Tawn:2012}
Wadsworth, J.~L. and Tawn, J.~A. (2012), \enquote{{Dependence Modelling for
  Spatial Extremes},} \textit{{Biometrika}}, 99, 253--272.

\bibitem[{Wadsworth and Tawn(2014)}]{wa14}
--- (2014), \enquote{Efficient Inference for Spatial Extreme Value Processes
  Associated to Log-Gaussian Random Functions,} \textit{Biometrika}, 101,
  1--15.

\end{thebibliography}

\end{document}